\documentclass[sigconf,natbib=true,screen=true]{acmart}

\acmSubmissionID{7389}

\usepackage{booktabs} % For formal tables
\usepackage{amsmath}
\usepackage{graphics}
\usepackage{epsfig}
\usepackage{graphicx}
\usepackage[ruled,vlined, linesnumbered]{algorithm2e}
\usepackage{xcolor}
\usepackage[skip=0pt]{caption}
\usepackage{subfigure}
\usepackage{balance}
\usepackage{multirow}
\usepackage{mathrsfs}
\usepackage{acronym}
\usepackage{placeins}
\usepackage{tabularx}
\usepackage{makecell}
\usepackage{xcolor}
\usepackage[inline]{enumitem}
\usepackage{fancyhdr}
\usepackage{soul}

\SetKwInput{KwInput}{Input} 
\SetKwInput{KwOutput}{Output}

\AtBeginDocument{%
  \providecommand\BibTeX{{%
    \normalfont B\kern-0.5em{\scshape i\kern-0.25em b}\kern-0.8em\TeX}}}

\newcommand{\changed}[1]{\textcolor{blue}{#1}}   

\acrodef{KL}{Kullback-Leibler}
\acrodef{BCE}{Binary Cross Entropy}
\acrodef{BPR}{bayesian personalized ranking}
\acrodef{GCN}{graph convolutional network}
\acrodef{GNN}{graph neural network}
\acrodef{CL}{contrastive learning}
\acrodef{VAE}{variational autoencoder}
\acrodef{KL-divergence}{Kullback–Leibler divergence}

\if0
\if0
\fi
\fi

\allowdisplaybreaks

\newcommand{\headernodot}[1]{\vspace*{1mm}\noindent\textbf{#1}}
\newcommand{\header}[1]{\headernodot{#1.}}

\author{Xin Xin*}
\orcid{0000-0001-6116-9115}
\affiliation{%
  \institution{Shandong University}
  \city{Qingdao}
  \country{China}
}
\email{xinxin@sdu.edu.cn}

\author{Xiangyuan Liu*}
\orcid{0009-0005-8513-0879}
\affiliation{%
  \institution{Shandong University}
  \city{Qingdao}
  \country{China}
}
\email{chrisxiangyuan@gmail.com}

\author{Hanbing Wang}
\orcid{0000-0002-5198-8085}
\affiliation{%
  \institution{Shandong University}
  \city{Qingdao}
  \country{China}
}
\email{hanbing.wang@mail.sdu.edu.cn}

\author{Pengjie Ren}
\orcid{0000-0003-2964-6422}
\affiliation{%
  \institution{Shandong University}
  \city{Qingdao}
  \country{China}
}
\email{jay.ren@outlook.com}

\author{Zhumin Chen}
\orcid{0000-0003-4592-4074}
\affiliation{%
  \institution{Shandong University}
  \city{Qingdao}
  \country{China}
}
\email{chenzhumin@sdu.edu.cn}

\author{Jiahuan Lei}
\orcid{0000-0002-5170-8645}
\affiliation{%
  \institution{Meituan}
  \city{Beijing}
  \country{China}
}
\email{leijiahuan@meituan.com}

\author{Xinlei Shi}
\orcid{0000-0002-0733-5757}
\affiliation{%
  \institution{Meituan}
  \city{Beijing}
  \country{China}
}
\email{shixinlei@meituan.com}

\author{Hengliang Luo}
\orcid{0000-0001-8597-8873}
\affiliation{%
  \institution{Meituan}
  \city{Beijing}
  \country{China}
}
\email{luohengliang@meituan.com}

\author{Joemon M. Jose}
\orcid{0000-0001-9228-1759}
\affiliation{%
  \institution{University of Glasgow}
  \city{London}
  \country{UK}
}
\email{Joemon.Jose@glasgow.ac.uk}

\author{Maarten de Rijke}
\orcid{0000-0002-1086-0202}
\affiliation{%
  \institution{University of Amsterdam}
  \city{Amsterdam}
  \country{The Netherlands}
}
\email{m.derijke@uva.nl}

\author{Zhaochun Ren$^{\dagger}$}
\orcid{0000-0002-9076-6565}
\affiliation{%
  \institution{Shandong University}
  \city{Qingdao}
  \country{China}
}
\email{zhaochun.ren@sdu.edu.cn}

\makeatletter
\def\authornotetext#1{
\if@ACM@anonymous\else
    \g@addto@macro\@authornotes{
    \stepcounter{footnote}\footnotetext{#1}}
\fi}
\makeatother
\authornotetext{Equal contribution.}
\authornotetext{Corresponding author.}

\def\authors{Xin Xin, Xiangyuan Liu, Hanbing Wang, Pengjie Ren, Zhumin Chen, Jiahuan Lei, Xinlei Shi, Hengliang Luo, Joemon M.Jose, Maarten de Rijke and Zhaochun Ren}
\renewcommand{\shortauthors}{Xin et al.}
\copyrightyear{2023}
\acmYear{2023}
\setcopyright{acmlicensed}\acmConference[SIGIR '23]{Proceedings of the 46th
International ACM SIGIR Conference on Research and Development in
Information Retrieval}{July 23--27, 2023}{Taipei, Taiwan}
\acmBooktitle{Proceedings of the 46th International ACM SIGIR Conference on
Research and Development in Information Retrieval (SIGIR '23), July 23--27,
2023, Taipei, Taiwan}
\acmPrice{15.00}
\acmDOI{10.1145/3539618.3591697}
\acmISBN{978-1-4503-9408-6/23/07}

\begin{document}
\begin{sloppypar}
% \fancyhead{} % use this for camera-ready, remove it when uploading to arxiv or your own website

\title[Improving  Recommender Systems through Multi-Behavior Alignment]{Improving Implicit Feedback-Based Recommendation \\through Multi-Behavior Alignment}    

\begin{abstract}
Recommender systems that learn from implicit feedback often use large volumes of a single type of implicit user feedback, such as clicks, to enhance the prediction of sparse target behavior such as purchases.
Using multiple types of implicit user feedback for such target behavior prediction purposes is still an open question.
Existing studies that attempted to learn from multiple types of user behavior often fail to:
\begin{enumerate*}[label=(\roman*)]
\item learn universal and accurate user preferences from different behavioral data distributions, and
\item overcome the noise and bias in observed implicit user feedback.
\end{enumerate*}

To address the above problems, we propose \textbf{m}ulti-\textbf{b}ehavior \textbf{a}lignment (MBA), a novel recommendation framework that learns from implicit feedback by using multiple types of behavioral data. 
We conjecture that multiple types of behavior from the same user (e.g., clicks and purchases) should reflect similar preferences of that user. 
To this end, we regard the underlying universal user preferences as a latent variable.
The variable is inferred by maximizing the likelihood of   multiple observed behavioral data distributions and, at the same time, minimizing the \acf{KL-divergence} between user models learned from auxiliary behavior (such as clicks or views) and the target behavior separately. 
MBA infers universal user preferences from multi-behavior data and performs data denoising to enable effective knowledge transfer. 
We conduct experiments on three datasets, including a dataset collected from an operational e-commerce platform. 
Empirical results demonstrate the effectiveness of our proposed method in utilizing multiple types of behavioral data to enhance the prediction of the target behavior.
% Code used to obtain the experimental results in this paper is available at \changed{\url{https://github.com/LiuXiangYuan/MBA}}.
\end{abstract}




\maketitle

%!TEX root = ../main.tex

\section{Introduction}

Recommender systems aim to infer user preferences from observed user-item interactions and recommend items that match those preferences. 
Many operational recommender systems are trained from implicit user feedback~\citep{gu2020hierarchical, huang2019online}.
Recommender systems that learn from implicit user feedback are typically trained on a single type of implicit user behavior, such as clicks.
However, in real-world scenarios, multiple types of user behavior are logged when a user interacts with a recommender system.
For example, users may click, add to a cart, and purchase items on an e-commerce platform~\citep{tsagkias-2020-challenges}.
Simply learning recommenders from a single type of behavioral data such as clicks can lead to a misunderstanding of a user's real user preferences since the click data is noisy and can easily be corrupted due to bias~\cite{chen2020bias}, and thus lead to suboptimal target behavior (e.g., purchases) predictions. 
Meanwhile, only considering purchase data tends to lead to severe cold-start problems~\citep{pan2019warm, xie2020internal, zhu2021transfer} and data sparsity problems~\citep{pan2010transfer, ma2022mixed}.

\header{Using multiple types of behavioral data}
How can we use multiple types of \emph{auxiliary} behavioral data (such as clicks) to enhance the prediction of sparse \emph{target} user behavior (such as purchases) and thereby improve recommendation performance? 
Some prior work~\citep{gao2019neural, chen2020efficient} has used multi-task learning to train recommender systems on both target behavior and multiple types of auxiliary behavior.
Building on recent advances in graph neural networks, \citet{jin2020multi} encode target behavior and multiple types of auxiliary behavior into a heterogeneous graph and perform convolution operations on the constructed graph for recommendation. 
% \citet{GuWSX22} focus on transferring information from different types of auxiliary behavior to target behavior through self-supervised learning on graphs. 
In addition, recent research tries to integrate the micro-behavior of user-item interactions into representation learning in the sequential and session-based recommendation~\citep{zhou2018micro, meng2020incorporating, yuan2022micro}.
These publications focus on mining user preferences from  user-item interactions, which is different from our task of predicting target behavior from multiple types of user behavior.

%\textcolor{red}{Xin - can you have a look at this para? we started with ref 2 and 11 but not explained it; then at the end we say our aim is "predicting target behavior from multiple types of auxiliary behavior"- in fact 13and 18 focuses on auxiliary behaviour- maybe we need to sharpen this part}
\begin{figure}
    \centering
    \includegraphics[clip,trim=2mm 0mm 0mm 0mm,width=0.45\textwidth]{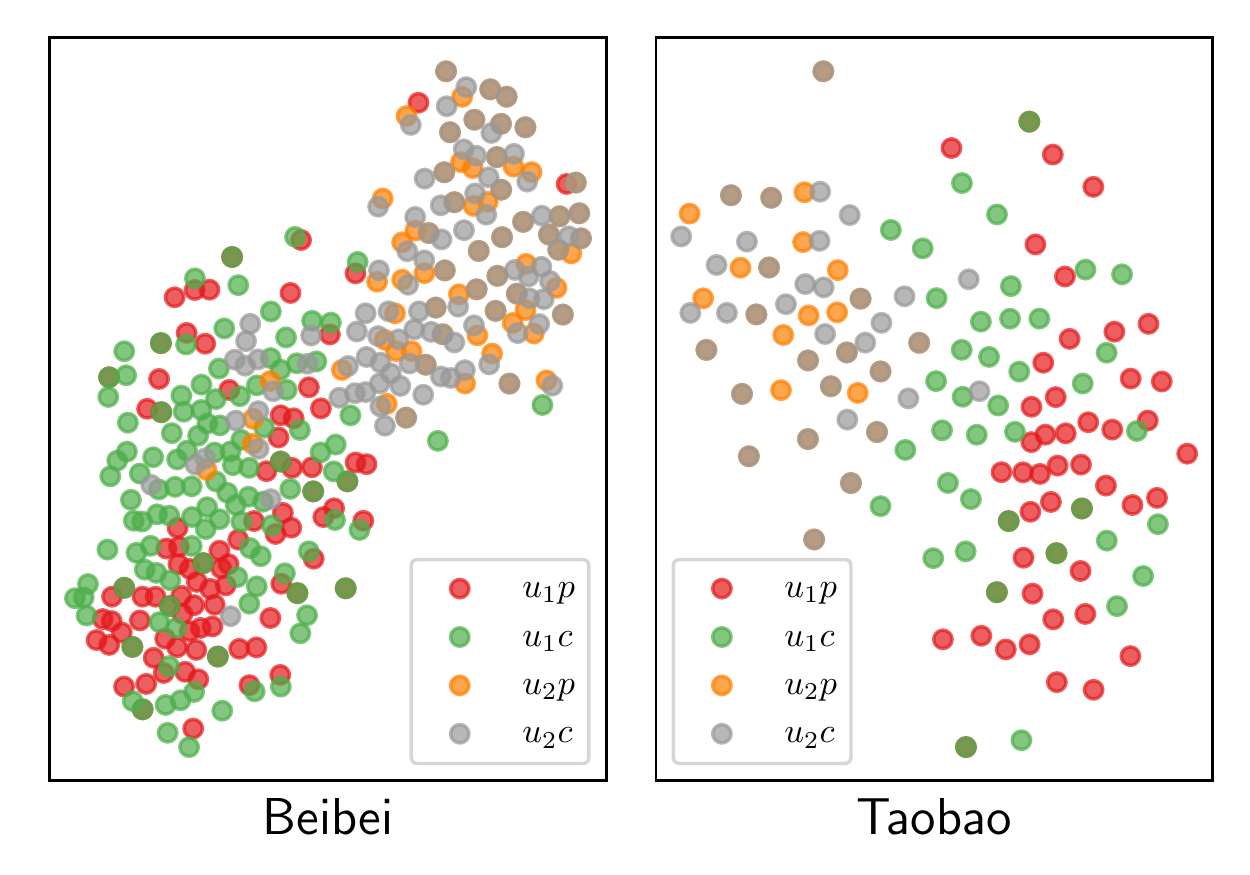}
    \caption{Distributions of items interacted with by two users in the Beibei and Taobao datasets (described in \S\ref{sec:data}). Item representations are obtained by a matrix factorization model trained on the purchase behavior data. $u_ic$ ($u_ip$) represents the distribution of items clicked (purchased) by user $u_i$.}
    \label{fig:pretrain_result}
    % \vspace{-0.1cm}
\end{figure}

\header{Limitations of current approaches}
Prior work on using multiple types of behavioral data to improve the prediction of the target behavior in a recommendation setting has two main limitations.

The first limitation concerns the gap between data distributions of different types of behavior. 
This gap impacts the learning of universal and effective user preferences.
For example, users may have clicked on but not purchased items, resulting in different positive and negative instance distributions across auxiliary and target behaviors.
Existing work typically learns separate user preferences for different types of behavior and then combines those preferences to obtain an aggregate user representation. We argue that:
\begin{enumerate*}[label=(\roman*)]
\item user preferences learned separately based on different types of behavior may not consistently lead to the true user preferences, and
\item multiple types of user behavior should reflect similar user preferences; in other words, there should be an underlying universal set of user preferences under different types of behavior of the same user.
\end{enumerate*}

The second limitation concerns the presence of noise and bias in auxiliary behavioral data, which impacts knowledge extraction and transfer.
A basic assumption of recommendations based on implicit feedback is that observed interactions between users and items reflect positive user preferences, while unobserved interactions are considered negative training instances. 
However, this assumption seldom holds in reality. 
A click may be triggered by popularity bias \citep{chen2020bias}, which does not reflect a positive preference. And an unobserved interaction may be attributed to a lack of exposure~\citep{chen2021bias}.
Hence, simply incorporating noisy or biased behavioral data may lead to sub-optimal recommendation performance.

\header{Motivation} 
Our assumption is that 
multiple types of behavior from the same user (e.g., clicks and purchases) should reflect similar preferences of that user.
To illustrate this assumption, consider Figure~\ref{fig:pretrain_result}, which shows distributions of items that two users ($u_1$ and $u_2$) interacted with (clicks $c$ and purchases $p$), in the Beibei and Taobao datasets (described in Section~\ref{sec:data} below). 
For both users, the items they clicked or purchased are relatively close. 
These observations motivate our hypothesis that multiple types of user behavior reflect similar user preferences, which is vital to improve the recommendation performance further.

\header{Proposed method}
To address the problem of learning from multiple types of auxiliary behavioral data and improve the prediction of the target behavior (and hence recommendation performance), we propose a training framework called \textbf{m}ulti-\textbf{b}ehavior \textbf{a}lignment (MBA). 
MBA aligns user preferences learned from different types of behavior.
The key assumption behind MBA is that multiple types of behavior from the same user reflect similar underlying user preferences.

To address the data distribution limitation mentioned above, we utilize \ac{KL-divergence} to measure the discrepancy between user models learned from multiple types of auxiliary behavior and target behavior, and then conduct knowledge transfer by minimizing this discrepancy to improve the recommendation performance.   

For the second limitation mentioned above (concerning noise and bias in behavioral data), MBA regards the underlying universal user preferences as a latent variable. The variable is then inferred by  maximizing the likelihood of multiple types of observed behavioral data while minimizing the discrepancy between models trained on different types of behavioral data. 
In this manner, MBA denoises multiple types of behavioral data and enables more effective knowledge transfer across multiple types of user behavior.

To demonstrate the effectiveness of the proposed method, we conduct extensive experiments on two open benchmark datasets and one dataset collected from an operational e-commerce platform. 
Experimental results show that the proposed MBA framework outperforms related state-of-the-art baselines.

\header{Main contributions}
Our main contributions are as follows:
\begin{itemize}[leftmargin=*,nosep]
    \item  We argue that multiple types of auxiliary and target behavior should reflect similar user preferences, and we propose to infer the true user preferences from multiple types of behavioral data.

    \item We propose a learning framework MBA to jointly perform data denoising and knowledge transfer across multiple types of behavioral data to enhance target behavior prediction and hence improve the recommendation performance.

    \item We conduct experiments on three datasets to demonstrate the effectiveness of the MBA method.
    One of these datasets is collected from an operational e-commerce platform, and includes clicks and purchase behavior data.
    Experimental results show state-of-the-art recommendation performance of the proposed MBA method.
\end{itemize}
%!TEX root = ../main.tex

\section{Related work}
We review prior work on multi-behavior recommendation and on denoising methods for recommendation from implicit feedback.

\subsection{Multi-behavior recommendation}
{Unlike conventional implicit feedback recommendation models~\citep{koren2009matrix, he2020lightgcn}, which train a recommender on a single type of user behavior (e.g., clicks), multi-behavior recommendation models use multiple types of auxiliary behavior data to enhance the recommendation performance on target behavior~\citep{gao2019neural, wei2022contrastive, xia2021graph, Cheng2023MBGCN, wang2020beyond, jin2020multi, chen2021graph}.
Recent studies use multi-task learning to perform joint optimization on learning auxiliary behavior and target behavior. 
For example, \citet{gao2019neural} propose a multi-task learning framework to learn user preferences from multi-behavior data based on a pre-defined relationship between different behavior.
Since different behavioral interactions between users and items can form a heterogeneous graph, recent studies also focus on using \acf{GNN} to mine the correlations among different types of behavior.
For example, \citet{wang2020beyond} uses the auxiliary behavior data to build global item-to-item relations and further improve the recommendation performance of target behavior.
\citet{jin2020multi} propose a \acf{GCN} based model on capturing the diverse influence of different types of behavior and the various semantics of different types of behavior.
\citet{xia2021graph} incorporate multi-behavior signals through graph-based meta-learning.
\citet{chen2021graph} regard the multi-behavior recommendation task as a multi-relationship prediction task and train the recommender with an efficient non-sampling method.
Additionally, some studies apply contrastive learning or a \acf{VAE} to improve the multi-behavior recommender.
% \citet{wei2022contrastive} propose a contrastive learning paradigm to capture the transferable user-item relationships from multi-behavior data. 
\citet{xuan2023knowledge} propose a knowledge graph enhanced contrastive learning framework to capture multi-behavioral dependencies better and solve the data sparsity problem of the target behavior, and~\citet{pan2022vae++} propose a \ac{VAE}-based model to conduct multi-behavior recommendation.

Another related research field is based on micro-behaviors~\citep{zhou2018micro, meng2020incorporating, yuan2022micro}, which utilize the micro-operation sequence in the process of user-item interactions to capture user preferences and predict the next item.
For example, \citet{yuan2022micro} focus on ``sequential patterns'' and ``dyadic relational patterns'' in micro-behaviors, and then use an extended self-attention network to mine the relationship between micro-behavior and user preferences. This work focuses on mining user preferences from the micro-operation sequence.}

However, existing studies still neglect the different data distributions across multiple types of user behavior, and thus fail to learn accurate and universal user preferences. Besides, prior work does not consider the noisy signals of user implicit feedback data, resulting in ineffective knowledge extraction and transfers. 

\subsection{Recommendation denoising}
{Existing recommender systems are usually trained with implicit feedback since it is much easier to collect than explicit ratings~\cite{rendle2012bpr}.
Recently, some research~\citep{jagerman2019model, wang2021denoising, wang2021implicit} has pointed out that implicit feedback can easily be  corrupted by different factors, such as various kinds of bias~\citep{chen2020bias} or users' mistaken clicks.
Therefore, there have been efforts aimed at alleviating the noisy problem of implicit recommendation. 
These efforts include sample selection methods~\citep{ding2018improved, ding2019reinforced, ding2019sampler, gantner2012personalized, yu2020sampler, wang2021implicit}, 
re-weighting methods~\citep{wang2021denoising, shen2019learning, wang2021denoising, wang2022learning, chen2022denoising}, 
methods using additional information~\citep{kim2014modeling, lu2019effects, zhao2016gaze}, 
and methods designing specific denoising architectures~\citep{chen2021structured, wu2021self, xie2021deep, gao2022self}.

Sample selection methods aim to design more effective samplers for model training.
For example, \citet{gantner2012personalized} consider popular but un-interacted items as items that are highly likely to be negative ones, while~\citet{ding2018improved} consider clicked but not purchased items as likely to be negative samples.
Re-weighting methods typically identify noisy samples as instances with higher loss values and then assign lower weights to them. For example, \citet{wang2021denoising} discard the large-loss samples with a dynamic threshold in each iteration.
\citet{wang2022learning} utilize the differences between model predictions as the denoising signals. 
Additional information such as dwell time~\citep{kim2014modeling}, gaze pattern~\citep{zhao2016gaze} and auxiliary item features~\citep{lu2019effects} can also be used to denoise implicit feedback.
Methods designing specific denoising architectures improve the robustness of recommender systems by designing special modules. 
\citet{wu2021self} use self-supervised learning on user-item interaction graphs to improve the robustness of graph-based recommendation models.
\citet{gao2022self} utilize the self-labeled memorized data as denoising signals to improve the robustness of recommendation models.}

Unlike the work listed above, which does not consider multiple types of user behavior, in this work, we focus on extracting underlying user preferences from (potentially) corrupted multi-behavior data and then conducting  knowledge transfer to improve the recommendation performance. 

%!TEX root = ../main.tex

\section{Method}

In this section, we detail our proposed MBA framework for multi-behavior recommendation.
We first introduce notations and the problem formulation in Section~\ref{sec:pf}.
After that, we describe how to perform multi-behavior alignment on noisy implicit feedback in Section~\ref{subsec:MBA}. Finally,  training details are given in Section~\ref{subsec:training}.

\subsection{Notation and problem formulation}
\label{sec:pf}
We write $u \in \mathcal{U}$ and $i \in \mathcal{I}$ for a user and an item, where $\mathcal{U}$ and $\mathcal{I}$ indicate the user set and the item set, respectively.
Without loss of generality, we regard click behavior as the auxiliary behavior and purchase behavior as the target behavior.
We write $\mathbf{R}_f \in \mathbb{R}^{|\mathcal{U}| \times |\mathcal{I}|}$ for the observed purchase behavior data.
Specifically, each item $r^{f}_{u,i} \in \mathbf{R}_f$ is set to 1 if there is a purchase behavior between user $u$ and item $i$; otherwise $r^{f}_{u,i}$ is set as 0.
Similarly, we denote $\mathbf{R}_g \in \mathbb{R}^{|\mathcal{U}| \times |\mathcal{I}|}$ as the observed click behavior data, where each $r^{g}_{u,i} \in \mathbf{R}_g$ is set as $1$ if there is a click behavior between user $u$ and item $i$; otherwise $r^{g}_{u,i}=0$.
We use $P(\mathbf{R}_f)$ and $P(\mathbf{R}_g)$ to denote the user preference distribution learned from $\mathbf{R}_f$ and $\mathbf{R}_g$, respectively.

{We assume that there is an underlying latent true user preference matrix $\mathbf{R}_t$ with $r^t_{u,i} \in \mathbf{R}_t$ as the true preference of user $u$ over item $i$. The probabilistic distribution of $\mathbf{R}_t$ is denoted as $P(\mathbf{R}_t)$.} 
Both the data observation of $\mathbf{R}_f$ and $\mathbf{R}_g$ is motivated by the latent universal true user preference distribution $P(\mathbf{R}_t)$ plus different kinds of noises or biases.
Formally, we assume that
%that $P(\mathbf{R}_f)$, $P(\mathbf{R}_g)$, and 
$P(\mathbf{R}_t)$ follows a Bernoulli distribution and can be approximated by a target recommender model $t_\theta$ with $\theta$ as the parameters:
\begin{equation}
    r^{t}_{u,i} \sim \operatorname{Bernoulli}(t_\theta(u,i)).
\end{equation}

\noindent%
Since the true user preferences $r^{t}_{u,i}$ are intractable, we need to introduce the learning signals from the observed $r^{f}_{u,i}$ and $r^{g}_{u,i}$ to infer $r^{t}_{u,i}$.  
% To better depict the latent true user preference, we assume that given the observed multi-behavior data, the latent true user preference $r^t_{u,i}$ is also drawn from Bernoulli distributions.
As a result, we introduce the following models to depict the correlations between the observed user implicit feedback (i.e., $r^{f}_{u,i}$ and $r^{g}_{u,i}$ ) and the latent true user preferences $r^{t}_{u,i}$:
\begin{equation}
\label{eq:correlations}
    \begin{split}
    &r^{f}_{u,i}\mid r^t_{u,i} = 0 \sim \operatorname{Bernoulli}(h^f_\phi(u,i)) \\
    &r^{f}_{u,i}\mid r^t_{u,i} = 1 \sim \operatorname{Bernoulli}(h^f_\varphi(u,i)) \\
    &r^{g}_{u,i}\mid r^t_{u,i} = 0 \sim \operatorname{Bernoulli}(h^g_{\phi '}(u,i)) \\
    &r^{g}_{u,i}\mid r^t_{u,i} = 1 \sim \operatorname{Bernoulli}(h^g_{\varphi '}(u,i)),
    \end{split}
\end{equation}
where \smash{$h^f_\phi(u,i)$} and \smash{$h^f_\varphi(u,i)$} are parameterized by $\phi$ and $\varphi$ in the observed purchase behavior data, respectively, while $h^g_{\phi '}(u,i)$ and $h^g_{\varphi '}(u,i)$ are parameterized by $\phi '$ and $\varphi '$ in the observed click behavior data respectively.

The target of our task is formulated as follows: given the observed multi-behavior user implicit feedback, i.e., $\mathbf{R}_f$ and $\mathbf{R}_g$, we aim to train the latent true user preference model $t_\theta$, and then use $t_\theta$ to improve the prediction performance on target behavior. {More precisely, during model inference, 
% because the test set is also part of the observed data, there is also noise, and what the target recommender $t_\theta$ represents the latent true user preference, which deviates from the distribution of the test set.
% So 
we introduce both $P(\mathbf{R_f})$ and $P(\mathbf{R_t})$ to perform the target behavior recommendation and use a hyperparameter $\beta$ to balance the $P(\mathbf{R_t})$ and $P(\mathbf{R_f})$, which is formulated as:
\begin{equation}
    \operatorname{score} = \beta P(\mathbf{R}_t) + (1 - \beta) P(\mathbf{R}_f).
\end{equation}
We select items with the highest score as the target behavior recommendation results.} 

\subsection{Multi-behavior alignment on noisy data}
\label{subsec:MBA}
The key motivation for MBA is that multiple types of user behavior should reflect similar user preferences. Hence,  Eq.~\ref{eq:similar} is expected to be achieved with the convergence of the training models:
\begin{equation} \label{eq:similar}
    P(\mathbf{R}_f) \approx P(\mathbf{R}_g) \approx P(\mathbf{R}_t).
\end{equation}
% 
% As described in section~\ref{sec:pf}, each $k$-th behavior user preference $P_k(\tilde{\mathbf{R}})$ can be represented as Eq.~\ref{eq:Pk}. 
% We use $P_a(\mathbf{R})$ and $P_t(\mathbf{R})$ to represent the auxiliary behavior real user preference and the target behavior real user preference, respectively.
% Ideally, both $P_a(\mathbf{R})$ and $P_t(\mathbf{R})$ approximate the universal user preference $P(\mathbf{R})$, which means
% \begin{equation} \label{eq:similar}
%     P_a(\mathbf{R}) \approx P_t(\mathbf{R}) \approx P(\mathbf{R})
% \end{equation}
Therefore, $P(\mathbf{R}_f)$ and $P(\mathbf{R}_t)$ should have a relatively small KL-divergence, which is formulated as follows:
% can transmit useful information from auxiliary behavior to target behavior, and we formulate it as
\begin{equation} \label{eq:kl}
\begin{aligned}
    KL[P(\mathbf{R}_f)\| P(\mathbf{R}_t)] = E_{P(\mathbf{R}_f)}[\log P(\mathbf{R}_f) - \log P(\mathbf{R}_t)].
\end{aligned}    
\end{equation}
Similarly, we also have the \ac{KL-divergence} between $P(\mathbf{R}_g)$ and $P(\mathbf{R}_t)$: 
\begin{equation} \label{eq:kl_aux}
\begin{aligned}
    KL[P(\mathbf{R}_g)\| P(\mathbf{R}_t)] = E_{P(\mathbf{R}_g)}[\log P(\mathbf{R}_g) - \log P(\mathbf{R}_t)].
\end{aligned}    
\end{equation}
However, naively minimizing the above \ac{KL-divergence} is not feasible since it overlooks the data distribution gaps and correlations between multiple types of behavior.
To address this issue, we use Bayes' theorem to rewrite $P(\mathbf{R}_t)$ as follows:
\begin{equation} \label{eq:bayes}
\begin{aligned}
    P(\mathbf{R}_t) = \frac{P(\mathbf{R}_f)P(\mathbf{R}_t\mid \mathbf{R}_f)}{P(\mathbf{R}_f\mid \mathbf{R}_t)} = \frac{P(\mathbf{R}_g)P(\mathbf{R}_t\mid \mathbf{R}_g)}{P(\mathbf{R}_g\mid \mathbf{R}_t)}.
\end{aligned}    
\end{equation}
By substituting the right part of Eq.~\ref{eq:bayes} into Eq.~\ref{eq:kl} and rearranging erms, we obtain the following equation: 
%, we rewrite the KL-diver\-gence $KL[P(\mathbf{R}_f)||P(\mathbf{R}_t)]$ as follows:
% \begin{equation} \label{eq:init_kl}
% \begin{aligned}
%     K&L[P(\mathbf{R}_f)||P(\mathbf{R}_t)] = \\
%     %E_{P(\mathbf{R}_f)}[\log P(\mathbf{R}_f) - \log P(\mathbf{R}_t)]
%     %&= E_{P(\mathbf{R}_f)}[\log P(\mathbf{R}_f) - \log \frac{P(\mathbf{R}_g)P(\mathbf{R}_t | \mathbf{R}_g)}{P(\mathbf{R}_g|\mathbf{R}_t)}]\\
%     %&= E_{P(\mathbf{R}_f)}[\log P(\mathbf{R}_f) - \log P(\mathbf{R}_g) - \log P(\mathbf{R}_t | \mathbf{R}_g) + \log P(\mathbf{R}_g | \mathbf{R}_t)]\\
%     & KL[P(\mathbf{R}_f)||P(\mathbf{R}_t|\mathbf{R}_g)] - \log P(\mathbf{R}_g) + E_{P(\mathbf{R}_f)}[\log P(\mathbf{R}_g|\mathbf{R}_t)].
% \end{aligned}    
% \end{equation}
%Then we rearrange Eq.~\ref{eq:init_kl} into the following equation:
\begin{equation} \label{eq:rearrange}
\begin{aligned}
    E_{P(\mathbf{R}_f)}[\log P & (\mathbf{R}_g\mid \mathbf{R}_t)] - KL[P(\mathbf{R}_f)\| P(\mathbf{R}_t)]\\
    &=\log P(\mathbf{R}_g) - KL[P(\mathbf{R}_f)\| P(\mathbf{R}_t\mid \mathbf{R}_g)].
\end{aligned}    
\end{equation}
Since $KL[P(\mathbf{R}_f)\| P(\mathbf{R}_t\mid \mathbf{R}_g)] \ge 0$, the left side of Eq.~\ref{eq:rearrange} is an approximate lower bound of the logarithm $\log P(\mathbf{R}_g)$.
The bound is satisfied if, and only if, $P(\mathbf{R}_f)$ perfectly recovers $P(\mathbf{R}_t\mid \mathbf{R}_g)$, which means $P(\mathbf{R}_f)$ trained on the observed target behavior can perfectly approximates the true user preference distribution captured from the auxiliary behavior data. The above condition is in line with the main motivation of the MBA, i.e., different behavior data should reflect similar user preferences.

% Similarly, if we substitute Eq.~\ref{eq:bayes} into Eq.~\ref{eq:kl_aux}, we can obtain:
We see that the left side of Eq.~\ref{eq:rearrange} is based on the expectation over $P(\mathbf{R}_f)$, which means that we are trying to train $P(\mathbf{R}_f)$ with the given corrupted auxiliary behavior data $\mathbf{R}_g$ (i.e., the term $E_{P(\mathbf{R}_f)}[\log P(\mathbf{R}_g\mid \mathbf{R}_t)]$) and then to transmit the information from $P(\mathbf{R}_f)$ to $P(\mathbf{R}_t)$ via the term $KL[P(\mathbf{R}_f)\|P(\mathbf{R}_t)]$.
Such a learning process is ineffective for learning the true user preference distribution $P(\mathbf{R}_t)$ and the target recommender model $t_\theta$.
%could build a gap among $P(\mathbf{R}_f)$, $P(\mathbf{R}_g)$ and $P(\mathbf{R}_t)$, since the observed data is corrupted and the three preferences are distributed in different spaces, which would affect the performance of our target recommender $t_\theta$.
To overcome the above issue, according to Eq.~\ref{eq:similar}, when the training process has converged, the preference distributions $P(\mathbf{R}_f)$ and $P(\mathbf{R}_t)$ would be close to each other.
As a result,  we can change the expectation over 
$P(\mathbf{R}_f)$ to the expectation over $P(\mathbf{R}_t)$ to learn $P(\mathbf{R}_t)$ more effectively. 
% $E_{P(\mathbf{R}_f)}[\log P(\mathbf{R}_g|\mathbf{R}_t)]$ over $P(\mathbf{R}_t)$ to align the true preference and purchase preference on noisy purchase data.
So we  modify the left side of Eq.~\ref{eq:rearrange} as
\begin{equation} \label{eq:rearrange_2}
\begin{split}
    E_{P(\mathbf{R}_t)}[\log P(&\mathbf{R}_g\mid\mathbf{R}_t)] - KL[P(\mathbf{R}_f)\|P(\mathbf{R}_t)]\\
    &\approx \log P(\mathbf{R}_g) - KL[P(\mathbf{R}_f)\|P(\mathbf{R}_t\mid\mathbf{R}_g)].
\end{split}    
\end{equation}
Similarly, if we substitute the middle part of Eq.~\ref{eq:bayes} into Eq.~\ref{eq:kl_aux} and perform similar derivations, we can obtain:
% Through similar derivation, we also have
\begin{equation} \label{eq:rearrange_t}
\begin{aligned}
    E_{P(\mathbf{R}_t)}[\log P(&\mathbf{R}_f\mid \mathbf{R}_t)] - KL[P(\mathbf{R}_g)\| P(\mathbf{R}_t)]\\
    &\approx \log P(\mathbf{R}_f) - KL[P(\mathbf{R}_g)\| P(\mathbf{R}_t\mid \mathbf{R}_f)].
\end{aligned}   
\end{equation}
The left side of Eq.~\ref{eq:rearrange_t} is an approximate lower bound of $\log P(\mathbf{R}_f)$.
The bound is satisfied only if $P(\mathbf{R}_g)$ perfectly recovers $P(\mathbf{R}_t\mid \mathbf{R}_f)$, which means $P(\mathbf{R}_g)$ trained on the observed auxiliary behaviors can perfectly approximate the true user preference distribution captured from the target behavior data. Such  condition further verifies the soundness of MBA, i.e., multiple types of user behavior are motivated by similar underlying user preferences.

Combining the left side of both Eq.~\ref{eq:rearrange_2} and Eq.~\ref{eq:rearrange_t}  we obtain the loss function as:
\begin{equation} \label{eq:loss}
\begin{aligned}
    L = & -E_{P(\mathbf{R}_t)}[\log P(\mathbf{R}_g\mid \mathbf{R}_t)] + KL[P(\mathbf{R}_f)\|P(\mathbf{R}_t)] \\
    & - E_{P(\mathbf{R}_t)}[\log P(\mathbf{R}_f\mid \mathbf{R}_t)] + KL[P(\mathbf{R}_g)\| P(\mathbf{R}_t)].
\end{aligned}
\end{equation}
We can see that the loss function aims to maximize the likelihood of data observation (i.e., $P(\mathbf{R}_g\mid \mathbf{R}_t)$ and $P(\mathbf{R}_f\mid \mathbf{R}_t)$) and minimize the \ac{KL-divergence} between distributions learned from different user behavior data.
The learning process
of MBA serves as a filter to simultaneously denoise multiple types of user behavior and conduct beneficial knowledge transfers to infer the true user preferences to enhance the prediction of the target behavior.
% In the training process, the target recommender will trained with the given corrupted multi-behaviors data and aligned between $P(\mathbf{R}_f)$ and $P(\mathbf{R}_g)$.
% According to the Eq.~\ref{eq:similar}, we only need to obtain one behavior real user preference to approximate the universal user preference. 
% Therefore, we focus on obtaining the target behavior real user preference $P_t(\mathbf{R})$.
% % with the help of the auxiliary behavior real user preference $P_a(\mathbf{R})$, 
% Since the term $E_{\mathbf{R}\sim P_a} [\log P(\tilde{\mathbf{R}}|\mathbf{R})]$ has no contribution to the training of $P_t(\mathbf{R})$, so we dump it and change our loss function as
% \begin{equation} \label{eq:loss}
%     \begin{aligned}
%         L = &\alpha KL[P_a(\mathbf{R})||P_t(\mathbf{R})] + (1 -\alpha)KL[P_t(\mathbf{R})||P_a(\mathbf{R})] \\
%         & - E_{\mathbf{R}\sim P_t}[\log P(\tilde{\mathbf{R}}|\mathbf{R})]
%     \end{aligned}
% \end{equation}

\subsection{Training details}
\label{subsec:training}

As described in Section~\ref{sec:pf}, we learn the user preference distributions $P(\mathbf{R}_f)$ and $P(\mathbf{R}_g)$ from $\mathbf{R}_f$ and $\mathbf{R}_g$, respectively.
In order to enhance the learning stability, we pre-train $P(\mathbf{R}_f)$ and $P(\mathbf{R}_g)$ in $\mathbf{R}_f$ and $\mathbf{R}_g$, respectively.
We use the same model structures of our target recommender $t_\theta$ as the pre-training model.

As the training converges, the \ac{KL-divergence} will gradually approach 0. 
In order to enhance the role of the \ac{KL-divergence} in conveying information, we set a hyperparameter $\alpha$ to enhance the effectiveness of the \ac{KL-divergence}.
Then we obtain the following training loss function:
\begin{equation} \label{eq:loss2}
    \begin{aligned}
    L_{MBA} &= -E_{P(\mathbf{R}_t)}[\log P(\mathbf{R}_g\mid \mathbf{R}_t)] + \alpha KL[P(\mathbf{R}_f)\|P(\mathbf{R}_t)] \\
    & - E_{P(\mathbf{R}_t)}[\log P(\mathbf{R}_f\mid \mathbf{R}_t)] + \alpha KL[P(\mathbf{R}_g)\|P(\mathbf{R}_t)].
    \end{aligned}
\end{equation}

\subsubsection{Expectation derivation}
As described in Section~\ref{sec:pf}, both $\mathbf{R}_f$ and $\mathbf{R}_g$ contain various kinds of noise and bias.
In order to infer the latent true user preferences from the corrupted multi-behavior data, we use $h^f_\phi(u,i)$ and $h^f_\varphi(u,i)$ to capture the correlations between the true user preferences and the observed purchase data. 
Similarly, $h^g_{\phi '}(u,i)$ and $h^g_{\varphi '}(u,i)$ are used to capture the correlations between the true user preferences and the observed click data, as shown in Eq.~\ref{eq:correlations}.
Specifically, 
%for a given true example (i.e., $r^t_{u,i} = 1$), we have both $r^f_{u,i} = 1$, $r^f_{u,i} = 0$ and $r^g_{u,i} = 1$, $r^g_{u,i} = 0$.
we expand $E_{P(\mathbf{R}_t)}[\log P(\mathbf{R}_g\mid \mathbf{R}_t)]$ as:
\begin{equation} \label{eq:Eg}
\begin{aligned}
    &E_{P(\mathbf{R}_t)}[\log P(\mathbf{R}_g\mid \mathbf{R}_t)] =  \sum_{(u,i)}E_{r^t_{u,i}\sim P(\mathbf{R}_t)}[\log P(r^g_{u,i} \mid r^t_{u,i})]\\
    % &= \sum_{(u,i)|r^g_{u,i} = 1} \left \{
    % \begin{aligned}
    %     &\log P(r^g_{u,i} = 1 | r^t_{u,i} = 1)\cdot P_t(r^t_{u,i} = 1)\\
    %     &+ \log P(r^g_{u,i} = 1 | r^t_{u,i} = 0)\cdot P_t(r^t_{u,i} = 0)
    % \end{aligned}
    % \right .\\
    % &+ \sum_{(u,i)|r^g_{u,i} = 0} \left \{
    % \begin{aligned}
    %     &\log P(r^g_{u,i} = 0 | r^t_{u,i} = 1)\cdot P_t(r^t_{u,i} = 1)\\
    %     &+ \log P(r^g_{u,i} = 0 | r^t_{u,i} = 0)\cdot P_t(r^t_{u,i} = 0)
    % \end{aligned}
    % \right .\\
    &=\sum_{(u,i)|r^g_{u,i} = 1}\left[\log h^g_{\varphi '}(u,i)t_\theta(u,i) +{}\right.\\
    &\hspace{1.75cm}\left.\log h^g_{\phi '}(u,i)(1 - t_\theta(u,i))\right] +{}\\
    &\phantom{=}
    \sum_{(u,i)|r^g_{u,i} = 0}\left[\log(1\!-\!h^g_{\varphi '}(u,i))t_\theta(u,i)+{}\right.\\
    &\hspace{1.6cm}\left.\log(1\!-\!h^g_{\phi '}(u,i))(1\!-\!t_\theta(u,i))\right].
\end{aligned}    
\end{equation}
Similarly, the term $E_{P(\mathbf{R}_t)}[\log P(\mathbf{R}_f\mid \mathbf{R}_t)]$ can be expanded as:
\begin{equation} \label{eq:Ef}
\begin{aligned}
    &E_{P(\mathbf{R}_t)}[\log P(\mathbf{R}_f\mid \mathbf{R}_t)] = \sum_{(u,i)}E_{r^t_{u,i}\sim P(\mathbf{R}_t)}[\log P(r^f_{u,i}\mid r^t_{u,i})]\\
    % &= \sum_{(u,i)|r^f_{u,i} = 1} \left \{
    % \begin{aligned}
    %     &\log P(r^f_{u,i} = 1 | r^t_{u,i} = 1)\cdot P_t(r^t_{u,i} = 1)\\
    %     &+ \log P(r^f_{u,i} = 1 | r^t_{u,i} = 0)\cdot P_t(r^t_{u,i} = 0)
    % \end{aligned}
    % \right .\\
    % &+ \sum_{(u,i)|r^f_{u,i} = 0} \left \{
    % \begin{aligned}
    %     &\log P(r^f_{u,i} = 0 | r^t_{u,i} = 1)\cdot P_t(r^t_{u,i} = 1)\\
    %     &+ \log P(r^f_{u,i} = 0 | r^t_{u,i} = 0)\cdot P_t(r^t_{u,i} = 0)
    % \end{aligned}
    % \right .\\
    &= \sum_{(u,i)|r^f_{u,i} = 1}\left[\log h^f_{\varphi}(u,i) t_\theta(u,i) \right.+{}\\[-3mm]
    &\hspace*{1.8cm}\left.\log h^f_{\phi}(u,i) (1 - t_\theta(u,i))\right] +{}\\
    &\mbox{}\hspace*{0.3cm} \sum_{(u,i)|r^f_{u,i} = 0} \left[\log (1 \!- \!h^f_{\varphi}(u,i))  t_\theta(u,i)\!\right. + {}\\[-3mm]
    &\mbox{}\hspace{1.8cm}
    \left.\log (1 \!- \!h^f_{\phi}(u,i)) (1 - t_\theta(u,i))\right].
\end{aligned}    
\end{equation}
% Then Eq.~\ref{eq:loss} can be rewrote as:
% \begin{equation} \label{eq:loss2}
%     \begin{aligned}
%     &L = KL[P(\mathbf{R}_f)||P(\mathbf{R}_t)] + KL[P(\mathbf{R}_g)||P(\mathbf{R}_t)] \\
%     & - \sum_{(u,i)|r^g_{u,i} = 1} \log h^g_{\varphi '}(u,i) \cdot t_\theta(u,i) + \log h^g_{\phi '}(u,i) \cdot (1 - t_\theta(u,i)) \\
%     & - \sum_{(u,i)|r^g_{u,i} = 0} \log (1 - h^g_{\varphi '}(u,i)) \cdot t_\theta(u,i) + \log (1 - h^g_{\phi '}(u,i)) \cdot (1 - t_\theta(u,i)) \\
%     & - \sum_{(u,i)|r^f_{u,i} = 1} \log h^f_{\varphi}(u,i) \cdot t_\theta(u,i) + \log h^f_{\phi}(u,i) \cdot (1 - t_\theta(u,i)) \\
%     & - \sum_{(u,i)|r^f_{u,i} = 0} \log (1 - h^f_{\varphi}(u,i)) \cdot t_\theta(u,i) + \log (1 - h^f_{\phi}(u,i)) \cdot (1 - t_\theta(u,i))
%     \end{aligned}
% \end{equation}
By aligning and denoising the observed target behavior and auxiliary behavior data simultaneously, the target recommender $t_\theta$ is trained to learn the universal true user preference distribution.

\subsubsection{Alternative model training} 
In the learning stage, we find that directly training $t_\theta$ with Eq.~\ref{eq:loss2}--Eq.~\ref{eq:Ef} does not yield satisfactory results, which is caused by the simultaneous update of five models (i.e., $h^g_{\phi '}$, $h^g_{\varphi '}$, $h^f_{\phi}$, $h^f_{\varphi}$ and $t_\theta$) in such an optimization process.
These five models may interfere with each other and prevent $t_\theta$ from learning well.
To address this problem, we set two alternative training steps to train the involved models iteratively.

In the first training step, we assume that
a user tends to not click or purchase items that the user dislikes. That is to say, 
% every negative example ($r^f_{u,i} = 0$ and $r^g_{u,i} = 0$) from multi-behavior data denotes 
% a true negative user preference.
given $r^t_{u,i} = 0$ we have $r^f_{u,i}\approx0$ and $r^g_{u,i}\approx0$, so we have $h^f_{\phi}\approx0$ and $h^g_{\phi '}\approx0$ according to Eq.~\ref{eq:correlations}.
Thus in this step, only the models $h^f_{\varphi}$,  $h^g_{\varphi '}$ and $t_\theta$ are trained.
Then Eq.~\ref{eq:Eg} can be reformulated as:
\begin{equation} \label{eq:EgC1}
    E_{P(\mathbf{R}_t)}[\log P(\mathbf{R}_g\mid \mathbf{R}_t)] = L_{CN} + L_{CP},
\end{equation}    
where
\begin{align*}
    L_{CN} & = \sum_{(u,i)\mid r^g_{u,i} = 0} \log (1 - h^g_{\varphi '}(u,i)) \cdot t_\theta(u,i),\\
    L_{CP} & = \sum_{(u,i)\mid r^g_{u,i} = 1} \log h^g_{\varphi '}(u,i) \cdot t_\theta(u,i) - C_1 \cdot (1 - t_\theta(u,i)).
\end{align*}
Meanwhile, Eq.~\ref{eq:Ef} can be reformulated as:
\begin{equation} \label{eq:EfC1}
    E_{P(\mathbf{R}_t)}[\log P(\mathbf{R}_f\mid \mathbf{R}_t)] = L_{PN} + L_{PP},
\end{equation}
where
\begin{align*}
    &L_{PN} = \sum_{(u,i)\mid r^f_{u,i} = 0} \log (1 - h^f_{\varphi}(u,i)) \cdot t_\theta(u,i),\\
    &L_{PP} = \sum_{(u,i)\mid r^f_{u,i} = 1} \log h^f_{\varphi}(u,i) \cdot t_\theta(u,i) - C_1 \cdot (1 - t_\theta(u,i)).
\end{align*}
% \begin{align} \label{eq:EgC1}
%     &L_{CN} = \sum_{(u,i)|r^g_{u,i} = 0} \log (1 - h^g_{\varphi '}(u,i)) \cdot t_\theta(u,i),\\
%     &L_{CP} = \sum_{(u,i)|r^g_{u,i} = 1} \log h^g_{\varphi '}(u,i) \cdot t_\theta(u,i) - C_1 \cdot (1 - t_\theta(u,i)),\\
%     &E_{P(\mathbf{R}_t)}[\log P(\mathbf{R}_g|\mathbf{R}_t)] = L_{CN} + L_{CP}
% \end{align}
% \begin{equation} \label{eq:EgC1}
%     \begin{aligned}
%     \small
%     &  E_{P(\mathbf{R}_t)}[\log P(\mathbf{R}_g|\mathbf{R}_t)] = 
%     \left . \!\!\!\!\!\!\!\! \sum_{(u,i)|r^g_{u,i} = 0} \!\!\!\!\!\!\!\! \log (1 - h^g_{\varphi '}(u,i)) \cdot t_\theta(u,i) \right\}{L_{CN}}   \\
%     &+ \left . \!\!\!\!\!\!\!\!  \sum_{(u,i)|r^g_{u,i} = 1} \!\!\!\!\!\!\!\! \log h^g_{\varphi '}(u,i) \cdot t_\theta(u,i) - C_1 \cdot (1 - t_\theta(u,i)) \right\}{L_{CP}}
%     \end{aligned}
% \end{equation}
% \begin{equation} \label{eq:EfC1}
%     \begin{aligned}
%     \small
%         & E_{P(\mathbf{R}_t)}[\log P(\mathbf{R}_f|\mathbf{R}_t)] = 
%     \left . \!\!\!\!\!\!\!\! \sum_{(u,i)|r^f_{u,i} = 0} \!\!\!\!\!\!\!\!\log (1 - h^f_{\varphi}(u,i)) \cdot t_\theta(u,i) \right\}{L_{PN}} \\
%     &+ \left . \!\!\!\!\!\!\!\! \sum_{(u,i)|r^f_{u,i} = 1} \!\!\!\!\!\!\!\! \log h^f_{\varphi}(u,i) \cdot t_\theta(u,i) - C_1 \cdot (1 - t_\theta(u,i)) \right\}{L_{PP}}
%     \end{aligned}
% \end{equation}
Here, we denote $C_1$ as a large positive hyperparameter to replace $-\log h^g_{\phi '}(u,i)$ and $-\log h^f_{\phi}(u,i)$.

In the second training step, we 
assume that a user tends to click and purchase the items that the user likes. That is to say, 
% regard every positive example ($r^f_{u,i}=1$ and $r^g_{u,i}=1$) from multi-behavior data as a true positive instance.
% Namely, 
given $r^t_{u,i} = 1$ we have $r^f_{u,i}\approx1$ and $r^g_{u,i}\approx1$, so we have $h^f_{\varphi}\approx1$ and $h^g_{\varphi '}\approx1$ according to Eq.~\ref{eq:correlations}.
Thus in this step, only the models $h^f_{\phi}$, $h^g_{\phi '}$ and $t_\theta$ will be updated.
Then Eq.~\ref{eq:Eg} can be reformulated as:
\begin{equation} \label{eq:EgC2}
    E_{P(\mathbf{R}_t)}[\log P(\mathbf{R}_g\mid \mathbf{R}_t)] = L'_{CP} + L'_{CN},
\end{equation}
where
\begin{align*}
    L'_{CP} & = \sum_{(u,i)\mid r^g_{u,i} = 1} \log h^g_{\phi '}(u,i) (1 - t_\theta(u,i)),\\
    L'_{CN} & =  \sum_{(u,i)\mid r^g_{u,i} = 0}  C_2  t_\theta(u,i) + \log (1 - h^g_{\phi '}(u,i))  (1 - t_\theta(u,i)).
\end{align*}
Eq.~\ref{eq:Ef} can be reformulated as:
\begin{equation} \label{eq:EfC2}
     E_{P(\mathbf{R}_t)}[\log P(\mathbf{R}_f\mid \mathbf{R}_t)] = L'_{PP} + L'_{PN}
\end{equation}     
where
\begin{align*}
    L'_{PP} & = \sum_{(u,i)\mid r^f_{u,i} = 1} \log h^f_{\phi}(u,i) (1 - t_\theta(u,i)),\\
    L'_{PN} & = \sum_{(u,i)\mid r^f_{u,i} = 0}  C_2  t_\theta(u,i) + \log (1 - h^f_{\phi}(u,i))  (1 - t_\theta(u,i)).
\end{align*}
% \begin{equation} \label{eq:EgC2}
%     \begin{aligned}
%     \small
%     & E_{P(\mathbf{R}_t)}[\log P(\mathbf{R}_g|\mathbf{R}_t)] =
%     \left . \!\!\!\!\!\!\!\! \sum_{(u,i)|r^g_{u,i} = 1} \!\!\!\!\!\!\!\! \log h^g_{\phi '}(u,i) \cdot (1 - t_\theta(u,i)) \right\}{L'_{CP}} \\
%     &- \left . \!\!\!\!\!\!\!\! \sum_{(u,i)|r^g_{u,i} = 0} \!\!\!\!\!\!\!\! C_2 \cdot t_\theta(u,i) + \log (1 - h^g_{\phi '}(u,i)) \cdot (1 - t_\theta(u,i)) \right\}{L'_{CN}}
%     \end{aligned}
% \end{equation}
% \begin{equation} \label{eq:EfC2}
%     \begin{aligned}
%     \small
%     & E_{P(\mathbf{R}_t)}[\log P(\mathbf{R}_f|\mathbf{R}_t)] = 
%     \left . \!\!\!\!\!\!\!\! \sum_{(u,i)|r^f_{u,i} = 1} \!\!\!\!\!\!\!\! \log h^f_{\phi}(u,i) \cdot (1 - t_\theta(u,i)) \right\}{L'_{PP}} \\
%     &- \left . \!\!\!\!\!\!\!\! \sum_{(u,i)|r^f_{u,i} = 0} \!\!\!\!\!\!\!\! C_2 \cdot t_\theta(u,i) + \log (1 - h^f_{\phi}(u,i)) \cdot (1 - t_\theta(u,i)) \right\}{L'_{PN}}
%     \end{aligned}
% \end{equation}
$C_2$ is a large positive hyperparameter to replace $-\log(1-h^g_{\varphi '}(u,i))$ and $-\log(1-h^f_{\varphi}(u,i))$.

\subsubsection{Training procedure} 
\label{sec:sample}
In order to facilitate the description of sampling and training process, we divide $E_{P(\mathbf{R}_t)}[\log P(\mathbf{R}_g\mid \mathbf{R}_t)]$ and $E_{P(\mathbf{R}_t)}[\log P(\mathbf{R}_f\mid \mathbf{R}_t)]$ into four parts (see Eq.~\ref{eq:EgC1} to Eq.~\ref{eq:EfC2}), namely \textbf{c}lick \textbf{p}ositive loss ($L_{CP}$ and $L'_{CP}$), \textbf{c}lick \textbf{n}egative loss ($L_{CN}$ and $L'_{CN}$), \textbf{p}urchase \textbf{p}ositive loss ($L_{PP}$ and $L'_{PP}$), and \textbf{p}urchase \textbf{n}egative loss ($L_{PN}$ and $L'_{PN}$).
Each sample in the training set can be categorized into one of  three situations: 
\begin{enumerate*}[label=(\roman*)]
\item clicked and purchased, 
\item clicked but not purchased, and 
\item not clicked and not purchased.
\end{enumerate*}
The three situations involve different terms in $E_{P(\mathbf{R}_t)}[\log P(\mathbf{R}_g\mid \mathbf{R}_t)]$ and $E_{P(\mathbf{R}_t)}[\log P(\mathbf{R}_f\mid \mathbf{R}_t)]$.
In situation (i), each sample involves the $L_{CP}$ and $L_{PP}$ (or $L'_{CP}$ and $L'_{PP}$ in the alternative training step).
In situation (ii), each sample involves the $L_{CP}$ and $L_{PN}$ (or $L'_{CP}$ and $L'_{PN}$ in the alternative training step).
In situation (iii), each sample involves the $L_{CN}$ and $L_{PN}$ (or $L'_{CN}$ and $L'_{PN}$ in the alternative training step).
% Each instance in the situation (iii) can be regarded as a negative instance, which involves the $L_{CN}$ and $L_{PN}$ (or $L'_{CN}$ and $L'_{PN}$ in the alternative training step).}
We then train MBA according to the observed multiple types of user behavior data in situations (i) and (ii), and use the samples in situation (iii) as our negative samples.
Details of the training process for MBA are provided in Algorithm~\ref{training}.

\begin{algorithm}[htb]
\SetAlgoLined
% \KwResult{Write here the result }
\KwInput{The observed multi-behavior data $\mathcal{D}$, hyperparameter settings;}
\KwOutput{All model parameters $\varphi$, $\varphi '$, $\phi$,$\phi'$,$\theta$;}
\While{not coverage}{
    Sample ($u$, $i$) from $\mathcal{D}$ \;
    flag = 0 \;
    $L_{KL} = \alpha KL[P(\mathbf{R}_f)\|P(\mathbf{R}_t)] + \alpha KL[P(\mathbf{R}_g)\|P(\mathbf{R}_t)]$ \;
    \uIf{flag=0}{
    \uIf{$r^f_{u,i} = 1$ and $r^g_{u,i} = 1$}{
            Compute $L_{MBA} = L_{KL} - (L_{CP} + L_{PP})$  \;
        }
    \uElseIf{$r^f_{u,i} = 0$ and $r^g_{u,i} = 1$}{
            Compute $L_{MBA} = L_{KL} - (L_{CP} + L_{PN})$  \;
    }
    \ElseIf{$r^f_{u,i} = 0$ and $r^g_{u,i} = 0$}{
            Compute $L_{MBA} = L_{KL} - (L_{CN} + L_{PN})$  \;
    }
    Update $\varphi$, $\varphi '$, and $\theta$ through $L_{MBA}$ \; 
    flag = 1 \;
    }
    \Else{
    Compute $L_{MBA}$ similar to line 6--line 12 using $L_{KL},L'_{PP},L'_{CP},L'_{PN},L'_{CN}$\;
        Update $\phi$, $\phi '$, and $\theta$ through $L_{MBA}$ \;
        flag = 0 \;
    }
}
 \caption{Training Process of MBA}
 \label{training}
\end{algorithm}

\section{Experimental settings}
\label{sec:expset}
%In this section, we detail our experimental methodology.

\subsection{Experimental questions}
Our experiments are conducted to answer the following research questions:
\begin{enumerate*}[label=\textbf{(RQ\arabic*)},leftmargin=*]
    \item How do the proposed methods perform compared with  state-of-the-art recommendation baselines on different datasets? 
    \item How do the proposed methods perform compared with other denoising frameworks?
    %How do different base models affect the recommendation performance?
    \item Can MBA help to learn universal user preferences from users' multiple types of behavior?
    % \item \textbf{RQ3.} Can the proposed methods also improve recommendation performance of the auxiliary click behavior?
    % \item \textbf{RQ3.} Can the proposed methods help to generate more robust recommendations on both click behavior data and purchase behavior data?
    % such as denoising and multi-behavior alignment?
    \item How do the components and the hyperparameter settings affect the recommendation performance of MBA?
\end{enumerate*}

\subsection{Datasets}
\label{sec:data}

To evaluate the effectiveness of our method, we conduct a series of experiments on three real-world benchmark datasets, including Beibei\footnote{\url{https://www.beibei.com/}} ~\cite{gao2019neural},
Taobao\footnote{\url{https://tianchi.aliyun.com/dataset/dataDetail?dataId=649}}~\cite{zhu2018learning}, and MBD (\textbf{m}ulti-\textbf{b}ehavior \textbf{d}ataset), a dataset we collected from an operational e-commerce platform. The details are as follows:
\begin{enumerate*}[label=(\roman*), leftmargin=*]
    \item The Beibei dataset is an open dataset collected from Beibei, the largest infant product e-commerce platform in China, which includes three types of behavior, \emph{click}, \emph{add-to-cart} and \emph{purchase}. This work uses two kinds of behavioral data, clicks and purchases.
    \item The Taobao dataset is an open dataset collected from Taobao, the largest e-commerce platform in China, which includes three types of behavior, \emph{click}, \emph{add to cart} and \emph{purchase}. In this work, we use clicks and purchases of this dataset.
    \item The MBD dataset is collected from an operational e-commerce platform, and includes two types of behavior, \emph{click} and \emph{purchase}. 
\end{enumerate*}
% \noindent%
For each dataset, we ensure that users have interactions on both types of behavior, and we set click data as auxiliary behavior data and purchase data as target behavior data.
Table~\ref{table1} shows the statistics of our datasets. 
% \begin{itemize}[leftmargin=*]
%     \item \textbf{Beibei.} There are three behaviors in this dataset, including click, add to cart and purchase, and purchase is the target behavior.
%     \item \textbf{Taobao.} There are three behaviors in this dataset, including click, add to cart and purchase, and purchase is the target behavior.
%     \item \textbf{MTBD.} There are two behaviors in this dataset, including click, add purchase, and purchase is the target behavior.
% \end{itemize}
% For Beibei and Taobao, we use the released data by MB-GMN~\cite{xia2021graph}, and Mt is collected from a real-world platform.

\begin{table}[t]
\caption{Statistics of the datasets.}
\label{table1}
\centering
    \begin{tabular}{l rrrr}
    \toprule
    Dataset & Users  & Items & Purchases & Clicks\\ 
    \midrule
    Beibei  & 21,716  & 7,977  & 243,661   & 1,930,069   \\
    Taobao  & 48,658  & 39,395 & 208,905   & 1,238,659   \\
    MBD    & 102,556 & 20,237 & 230,958   & 659,914    \\ 
    \bottomrule
    \end{tabular}
\end{table}

\subsection{Evaluation protocols}
We divide the datasets into training and test sets with a ratio of  4:1.
We adopt two widely used metrics Recall@$k$ and NDCG@$k$. 
Recall@$k$ represents the coverage of true positive items that appear in the final top-$k$ ranked list. 
NDCG@$k$ measures the ranking quality of the final recommended items.
In our experiments, we use the setting of $k=10,20$.
For our method and the baselines, the reported results are the average values over all users.
For every result, we conduct the experiments three times and report the average values.

\subsection{Baselines}
To demonstrate the effectiveness of our method, we compare MBA with several state-of-the-art methods.
The methods used for comparison include single-behavior models, multi-behavior models, and recommendation denoising methods.

The single-behavior models that we consider are:
\begin{enumerate*}[label=(\roman*), leftmargin=*]
    \item \textbf{MF-BPR}~\cite{rendle2012bpr}, which uses \acf{BPR} loss to optimize matrix factorization.
    \item \textbf{NGCF}~\cite{wang2019neural}, which encodes collaborative signals into the embedding process through multiple graph convolutional layers and models higher-order connectivity in user-item graphs.
    \item \textbf{LightGCN}~\cite{he2020lightgcn}, which simplifies graph convolution by removing the matrix transformation and non-linear activation. We use the \ac{BPR} loss to optimize LightGCN.
\end{enumerate*}

% \noindent%
The multi-behavior models that we consider are:
\begin{enumerate*}[label=(\roman*), leftmargin=*]
% \item \textbf{BPR}~\cite{rendle2012bpr}: BPR is a popular matrix factorization mode trained with Bayesian personalized ranking loss on a single behavior.

\item \textbf{MB-GCN}~\cite{jin2020multi}, which constructs a multi-behavior heterogeneous graph and uses \ac{GCN} to perform behavior-aware embedding propagation.

\item \textbf{MB- GMN}~\cite{xia2021graph}, which incorporates multi-behavior pattern modeling with the meta-learning paradigm.

\item \textbf{CML}~\cite{wei2022contrastive}, which uses a new multi-behavior contrastive learning paradigm to capture the transferable user-item relationships from multi-behavior data.

\end{enumerate*}

% \noindent%
To verify that the proposed method improves performance by denoising implicit feedback, we also introduce the following denoising frameworks:
\begin{enumerate*}[label=(\roman*), leftmargin=*]

\item \textbf{WBPR}~\cite{gantner2012personalized}, which is a re-sampling-based method which considers popular, but un-interacted items are highly likely to be negative. 

\item \textbf{T-CE}~\cite{wang2021denoising}, which is a re-weighting based method which discards the large-loss samples with a dynamic threshold in each iteration.

\item \textbf{DeCA}~\cite{wang2022learning}, which is a newly proposed denoising method that utilizes the agreement predictions on clean examples across different models and minimizes the \ac{KL-divergence} between the real user preference parameterized by two recommendation models.  

\item \textbf{SGDL}~\cite{gao2022self}, which is a new denoising paradigm that utilizes self-labeled memorized data as denoising signals to improve the robustness of recommendation models.

\end{enumerate*}
\begin{table*}[h]
\caption{Overall performance comparison of purchase predictions on Beibei, Taobao and MBD with single-behavior methods and multi-behavior methods.
R denotes Recall, and N denotes NDCG.
Underscore indicates the best result among the baseline methods. Boldface means best results.
A significant improvement over the best baseline is marked with * ($p<0.05$).}
\label{tb:multi_result}
\centering
    \resizebox{\linewidth}{!}{
        \begin{tabular}{l cccc cccc cccc}
    \toprule
    \textbf{Datasets} & \multicolumn{4}{c}{\textbf{Beibei}} & \multicolumn{4}{c}{\textbf{Taobao}} & \multicolumn{4}{c}{\textbf{MBD}}                               \\ 
    \cmidrule(r){2-5}\cmidrule(r){6-9}\cmidrule{10-13}
    Method            & R@10   & R@20   & N@10   & N@20   & R@10   & R@20   & N@10   & N@20   & R@10   & R@20   & N@10   & N@20   \\ 
    \midrule
    MF                & 0.0901 & 0.1438 & 0.0668 & 0.0855 & 0.0303 & 0.0420 & 0.0190 & 0.0224 & 0.3816 & 0.4597 & 0.2490 & 0.2697 \\
    NGCF                & 0.0987 & 0.1561 & 0.0742 & 0.0939 & 0.0350 & 0.0518 & 0.0219 & 0.0267 & 0.4030 & 0.4892 & 0.2623 & 0.2852 \\
    LightGCN          & 0.0988 & 0.1541 & 0.0733 & 0.0925 & 0.0460 & 0.0640 & 0.0290 & 0.0342 & 0.3460 & 0.4365 & 0.2205 & 0.2443 \\ \midrule
    MBGCN             & \ul{0.1054} & \ul{0.1642} & \ul{0.0784} & \ul{0.0986} & 0.0418 & 0.0621 & 0.0250 & 0.0308 & \ul{0.4625} & \ul{0.5615} & \ul{0.2949} & \ul{0.3213} \\
    MBGMN             & 0.1046 & 0.1632 & 0.0779 & 0.0981 & \ul{0.0542} & \ul{0.0808} & \ul{0.0330} & \ul{0.0406} & 0.3660 & 0.4810 & 0.2323 & 0.2624 \\
    CML               & 0.0861 & 0.1382 & 0.0618 & 0.0796 & 0.0318 & 0.0524 & 0.0176 & 0.0234 & 0.3839 & 0.4928 & 0.2200 & 0.2487 \\ \midrule
    MBA           & \textbf{0.1127} & \textbf{0.1742}\rlap{*} & \textbf{0.0834} & \textbf{0.1046}\rlap{*} & \textbf{0.0579}\rlap{*} & \textbf{0.0812} & \textbf{0.0369}\rlap{*} & \textbf{0.0435}\rlap{*} & \textbf{0.4644} & \textbf{0.5677}\rlap{*} & \textbf{0.3012}\rlap{*} & \textbf{0.3285}\rlap{*} \\
    \bottomrule
    \end{tabular}
    }
%\vspace{-0.3cm}
\end{table*}
\begin{table*}
% \vspace{-0.25cm}
\caption{Overall performance comparison of purchase predictions on Beibei, Taobao and MBD with other denoising frameworks. Normal denotes the normal training. R denotes Recall, and N denotes NDCG. Underscore indicates the best result among the baseline methods. Boldface indicates best results.
A significant improvement over the best baseline is marked with * ($p<0.05$).
% The results with improvements over the best baseline larger than 5\% are marked with $*$ .
}
\setlength{\tabcolsep}{1.25mm}
%Significant improvement over the best baseline is marked with * ($p<0.05$).}
\label{tb:denoise_purchase}
\centering
    \resizebox{\linewidth}{!}{
        \begin{tabular}{l @{ \ }l @{ \ } cccc cccc cccc}
            \toprule
            \multicolumn{2}{l}{\textbf{Dataset}}                  & \multicolumn{4}{c}{\textbf{Beibei}}                   & \multicolumn{4}{c}{\textbf{Taobao}}                   & \multicolumn{4}{c}{\textbf{MBD}}                               \\ \midrule
            Base model & Method & R@10   & R@20   & N@10   & N@20   & R@10   & R@20   & N@10   & N@20   & R@10   & R@20   & N@10   & N@20   \\ 
            \midrule
            \multicolumn{1}{l}{\multirow{6}{*}{MF}}       & Normal & \ul{0.0901} & \ul{0.1438} & \ul{0.0668} & \ul{0.0855} & 0.0303 & 0.0420 & 0.0190 & 0.0224 & \ul{0.3816} & \ul{0.4597} & \ul{0.2490} & \ul{0.2697} \\
             & WBPR   & 0.0770 & 0.1189 & 0.0586 & 0.0731 & \ul{0.0307} & \ul{0.0440} & \ul{0.0195} & \ul{0.0233} & 0.3554 & 0.4216 & 0.2375 & 0.2551 \\
             & T-CE   & 0.0795 & 0.1262 & 0.0562 & 0.0724 & 0.0261 & 0.0350 & 0.0148 & 0.0174 & 0.1868 & 0.2320 & 0.1195 & 0.1314 \\
             & DeCA   & 0.0898 & 0.1378 & 0.0664 & 0.0831 & 0.0282 & 0.0405 & 0.0183 & 0.0218 & 0.3055 & 0.3654 & 0.1983 & 0.2143 \\
             & SGDL   & 0.0768 & 0.1240 & 0.0561 & 0.0725 & 0.0090 & 0.0137 & 0.0056 & 0.0069 & 0.1833 & 0.2300 & 0.1183 & 0.1307 \\ 
             \cmidrule{2-14} 
             & MBA    & \textbf{0.1127}\rlap{*} & \textbf{0.1742}\rlap{*} & \textbf{0.0834}\rlap{*} & \textbf{0.1046}\rlap{*} & \textbf{0.0579}\rlap{*} & \textbf{0.0812}\rlap{*} & \textbf{0.0369}\rlap{*} & \textbf{0.0435}\rlap{*} & \textbf{0.4644}\rlap{*} & \textbf{0.5677}\rlap{*} & \textbf{0.3012}\rlap{*} & \textbf{0.3285}\rlap{*} \\ 
            \midrule 
            \multirow{6}{*}{LightGCN} & Normal & \ul{0.0988} & \ul{0.1541} & \ul{0.0733} & \ul{0.0925} & 0.0460 & 0.0640 & 0.0290 & 0.0342 & \ul{0.4145} & \ul{0.4974} & \ul{0.2719} & \ul{0.2939} \\
             & WBPR   & 0.0879 & 0.1310 & 0.0682 & 0.0832 & \ul{0.0486} & \ul{0.0694} & \ul{0.0307} & \ul{0.0367} & 0.3989 & 0.4772 & 0.2647 & 0.2855 \\
             & T-CE   & 0.0779 & 0.1254 & 0.0526 & 0.0691 & 0.0292 & 0.0416 & 0.0188 & 0.0224 & 0.2761 & 0.3465 & 0.1756 & 0.1941 \\
             & DeCA   & 0.0770 & 0.1230 & 0.0528 & 0.0687 & 0.0412 & 0.0587 & 0.0255 & 0.0305 & 0.2382 & 0.3031 & 0.1544 & 0.1714 \\
             & SGDL   & 0.0944 & 0.1462 & 0.0695 & 0.0874 & 0.0452 & 0.0641 & 0.0281 & 0.0335 & 0.3894 & 0.4707 & 0.2470 & 0.2686 \\ 
             \cmidrule{2-14} 
            & MBA    & \textbf{0.1078}\rlap{*} & \textbf{0.1698}\rlap{*} & \textbf{0.0786}\rlap{*} & \textbf{0.1002}\rlap{*} & \textbf{0.0538}\rlap{*} & \textbf{0.0780}\rlap{*} & \textbf{0.0327}\rlap{*} & \textbf{0.0396}\rlap{*} & \textbf{0.4750}\rlap{*} & \textbf{0.5926}\rlap{*} & \textbf{0.3026}\rlap{*} & \textbf{0.3337}\rlap{*} \\ 
            \bottomrule
        \end{tabular}
    }
%\vspace{-0.3cm}
\end{table*}
\subsection{Implementation details}
We implement our method with PyTorch.\footnote{Our code is available at \changed{\url{https://github.com/LiuXiangYuan/MBA}}.} Without special mention, we set MF as our base model $t_\theta$ since MF is still one of the best models for capturing user preferences for recommendations~\cite{rendle2020neural}.
The model is optimized by Adam~\cite{kingma2014adam} optimizer with a learning rate of 0.001, where the batch size is set as 2048. 
The embedding size is set to 32.
The hyperparameters $\alpha$, $C_1$ and $C_2$ are search from \{ 1, 10, 100, 1000 \}.
$\beta$ is search from \{ 0.7, 0.8, 1 \}.
To avoid over-fitting, $L_2$ normalization is searched in \{ $10^{-6}$, $10^{-5}$, \ldots, 1 \}. 
Each training step is formed by one interacted example, and one randomly sampled negative example for efficient computation.
We use Recall@20 on the test set for early stopping if the value does not increase after 20 epochs.

For the hyperparameters of all recommendation baselines, we use the values suggested by the original papers with carefully fine-tuning on the three datasets.
For all graph-based methods, the number of graph-based message propagation layers is fixed at 3.

%!TEX root = ../main.tex

\section{Experimental Results}

% We present our experimental results and answer our research questions.

\subsection{Performance comparison (RQ1)}

To answer $\rm RQ1$, we conduct experiments on the Beibei, Taobao and MBD datasets.
The performance comparisons are reported in Table~\ref{tb:multi_result}.
From the table, we have the following observations.

First, the proposed MBA method achieves the best performance and consistently outperforms all baselines across all datasets.
For instance, the average improvement of MBA over the strongest baseline is approximately 6.3\% on the Beibei dataset, 6.6\% on the Taobao dataset and 1.5\% on the MBD dataset. 
These improvements demonstrate the effectiveness of MBA.
We contribute the significant performance improvement to the following two reasons:
(i) we align the user preferences based on two types of two behavior, transferring useful information from the auxiliary behavior data to enhance the performance of the target behavior predictions;
(ii) noisy interactions are reduced through preference alignment, which helps to improve the learning of the latent universal true user preferences.

Second, except CML the multi-behavior models outperform the single-behavior models by a large margin.
This reflects the fact that adding auxiliary behavior information can improve the recommendation performance of the target behavior.
We conjecture that CML cannot achieve satisfactory performance because it incorporates the knowledge contained in auxiliary behavior through contrastive meta-learning, which introduces more noisy signals.
Furthermore, we compare MBA with the best single-behavior model (NGCF on  the Beibei and MBD datasets, LightGCN on the Taobao dataset), and see that MBA achieves an average improvement of 12.4\% on the Beibei dataset, 26.8\% on the Taobao dataset and 15.3\% on the MBD dataset.

To conclude, the proposed MBA approach consistently and significantly outperforms related single-behavior and multi-behavior recommendation baselines on the purchase prediction task.

\subsection{Denoising (RQ2)}
Table~\ref{tb:denoise_purchase} reports on a performance comparison with existing denoising frameworks on the Beibei, Taobao and MBD datasets. 
The results demonstrate that MBA can provide more robust recommendations and improve overall performance than competing approaches.
Most of the denoising baselines do not obtain satisfactory results, even after carefully tuning their hyperparameters.
Only WBPR can outperform normal training in some cases. However, MBA consistently outperforms normal training and other denoising frameworks.
We think the reasons for this are as follows: 
(i) In MBA, we use the alignment between multi-behavior data as the denoising signal and then transmit information from the multi-behavior distribution to the latent universal true user preference distribution. This learning process facilitates knowledge transfer across multiple types of user behavior and filters out noisy signals.
(ii) In the original papers of the compared denoising baselines, testing is conducted based on explicit user-item ratings. However, our method does not use any explicit information like ratings, only implicit interaction data is considered.

To further explore the generalization capability of MBA, we also adopt LightGCN as our base model (i.e., using LightGCN as $t_\theta$).
The results are also shown in Table~\ref{tb:denoise_purchase}.
We see that MBA is still more effective than the baseline methods.
We find that LightGCN-based MBA does not perform as well as MF-based MBA on the Beibei and Taobao datasets.
We think the possible reasons are as follows: 
(i) LightGCN is more complex than MF, making MBA more difficult to train;
(ii) LightGCN may be more sensitive to noisy signals due to the aggregation of neighbourhoods, resulting in a decline in the MBA performance compared to using MF as the base model.

To conclude, the proposed MBA can generate more accurate recommendation compared with existing denoising frameworks.
\vspace{-1mm}
\subsection{User preferences visualization (RQ3)}

\begin{figure}[t]
    %\vspace{-0.5cm}
    % \setlength{\belowcaptionskip}{-0.3cm}
    \centering
    \includegraphics[clip,trim=2mm 0mm 0mm 0mm,width=\linewidth]{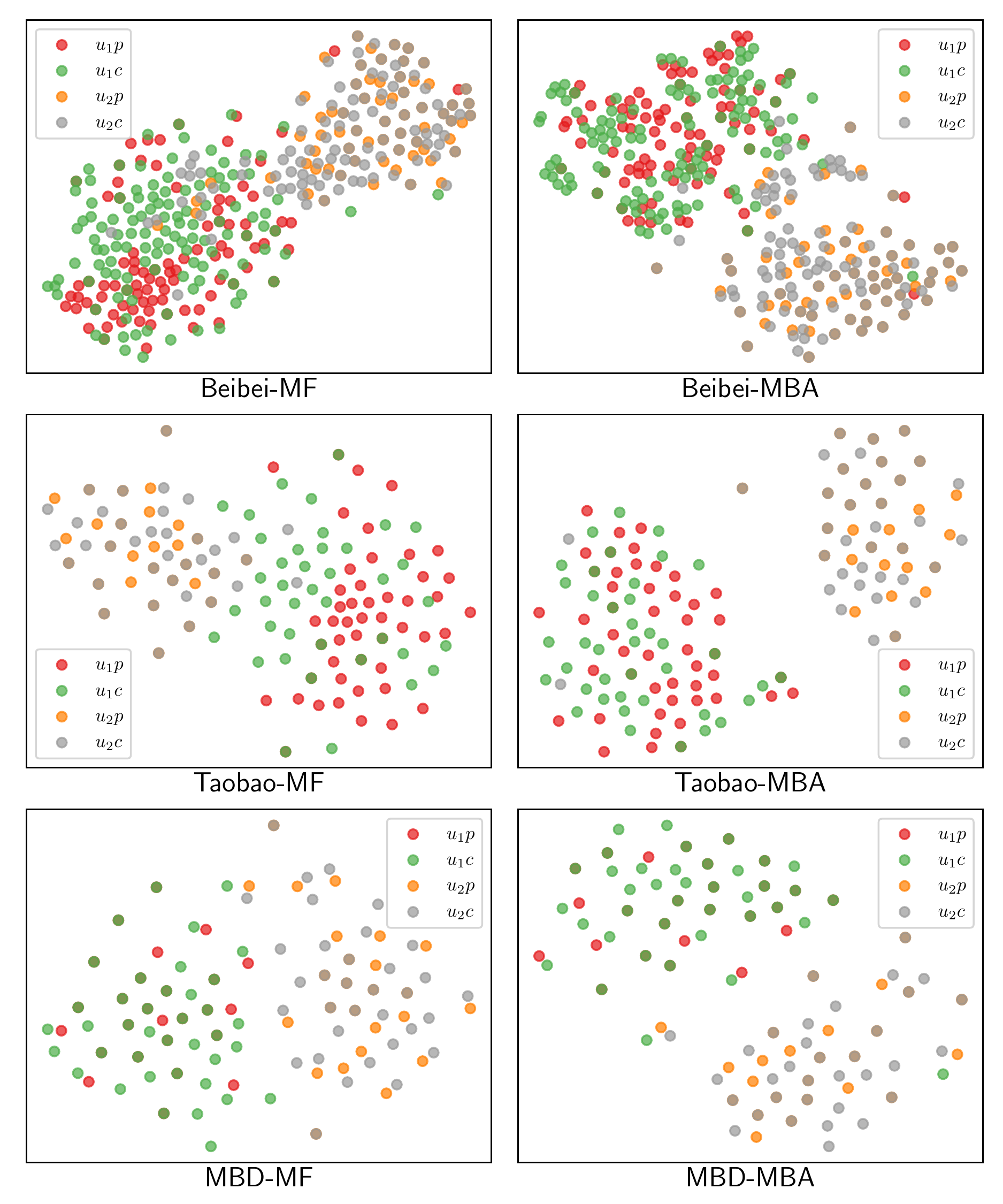}
    \caption{
    Distributions of items interacted with by two users in the Beibei, Taobao and MBD datasets.
    % Two users' behavior distributions in Beibei dataset, Taobao dataset and MBD dataset. 
    Item representations in the graphs on the left are obtained by a matrix factorization model trained on the purchase behavior data, while item representations in the right column graphs are obtained by MBA.
    $u_ic$ ($u_ip$) represents the distribution of items clicked (purchased) by user $u_i$.
    }
    \label{fig:target_result}
\end{figure}

To answer RQ3, we visualize the distribution of users' interacted items.
We select two users in the Beibei, Taobao and MBD datasets and draw their behavior distributions using the parameters obtained from an MF model trained on the purchase behavior data and the parameters obtained from MBA, respectively.
Figure~\ref{fig:target_result} visualizes the results. 
From the figure, we observe that for one user, the clicked items and purchased items distributions of MBA stay much closer than that of MF.
The observation indicates that MBA can successfully align multiple types of user behavior and infer universal and accurate user preferences. 

Besides, we see that different users in MBA have more obvious separations than users in MF, which implies that MBA provides more personalized user-specific recommendation than MF.

\vspace{-4mm}
\subsection{Model investigation (RQ4)}

\begin{table*}
\caption{The performance of MBA after removing \ac{KL-divergence} and pre-trained models. R denotes Recall, and N denotes NDCG. Boldface indicates best results. A significant improvement over the best baseline is marked with * ($p<0.05$)}
\label{tb:ablation_result}
\centering
%    \resizebox{\linewidth}{!}{
        \begin{tabular}{l cccc cccc cccc}
    \toprule
    \textbf{Datasets} & \multicolumn{4}{c}{\textbf{Beibei}}                   & \multicolumn{4}{c}{\textbf{Taobao}}                   & \multicolumn{4}{c}{\textbf{MBD}}                               \\ 
    \cmidrule(r){2-5}\cmidrule(r){6-9}\cmidrule{10-13}
    Method            & R@10   & R@20   & N@10   & N@20   & R@10   & R@20   & N@10   & N@20   & R@10   & R@20   & N@10   & N@20   \\ 
    \midrule
    MBA-KL                & 0.0897 & 0.1412 & 0.0651 & 0.0831 & 0.0261 & 0.0380 & 0.0151 & 0.0185 & 0.3250 & 0.4185 & 0.1779 & 0.2027 \\
    MBA-PT                & 0.0687 & 0.1136 & 0.0487 & 0.0642 & 0.0087 & 0.0152 & 0.0054 & 0.0072 & 0.3226 & 0.4138 & 0.1775 & 0.2017 \\
    MBA           & \textbf{0.1127}\rlap{*} & \textbf{0.1742}\rlap{*} & \textbf{0.0834}\rlap{*} & \textbf{0.1046}\rlap{*} & \textbf{0.0579}\rlap{*} & \textbf{0.0812}\rlap{*} & \textbf{0.0369}\rlap{*} & \textbf{0.0435}\rlap{*} & \textbf{0.4644}\rlap{*} & \textbf{0.5677}\rlap{*} & \textbf{0.3012}\rlap{*} & \textbf{0.3285}\rlap{*} \\ 
    \bottomrule
    \end{tabular}
%    }
\end{table*}

\subsubsection{Ablation study.}
Regarding RQ4, we conduct an ablation study (see Table~\ref{tb:ablation_result}) on the following two settings:
\begin{enumerate*}[label=(\roman*), leftmargin=*] 
\item \textbf{MBA-KL}: we remove \ac{KL-divergence} when training MBA; and
\item \textbf{MBA-PT}: we co-train the $P(\mathbf{R}_f)$ and $P(\mathbf{R}_g)$ in MBA instead of pre-training.
\end{enumerate*}

The results show that both parts (\ac{KL-divergence} and pre-trained models) are essential to MBA because removing either will lead to a performance decrease.
Without \ac{KL-divergence}, we see the performance drops substantially in terms of all metrics. %e.g., 19\%, 53\% and 26\% in Recall@20 on the Beibei, Taobao and MBD datasets, respectively.
Hence, the \ac{KL-divergence} helps align the user preferences learned from different behaviors, thus improving the recommendation performance.
Without pre-trained models, the results drop dramatically, especially in the Taobao dataset, which indicates that it is hard to co-train $P(\mathbf{R}_f)$ and $P(\mathbf{R}_g)$ with MBA.
{Using a pre-trained model can reduce MBA's complexity and provide prior knowledge so that it can more effectively extract the user's real preferences from the different types of behavior distributions.}

\subsubsection{Hyperparameter study}

\begin{figure}[t]
    \centering
    \includegraphics[width=\linewidth]{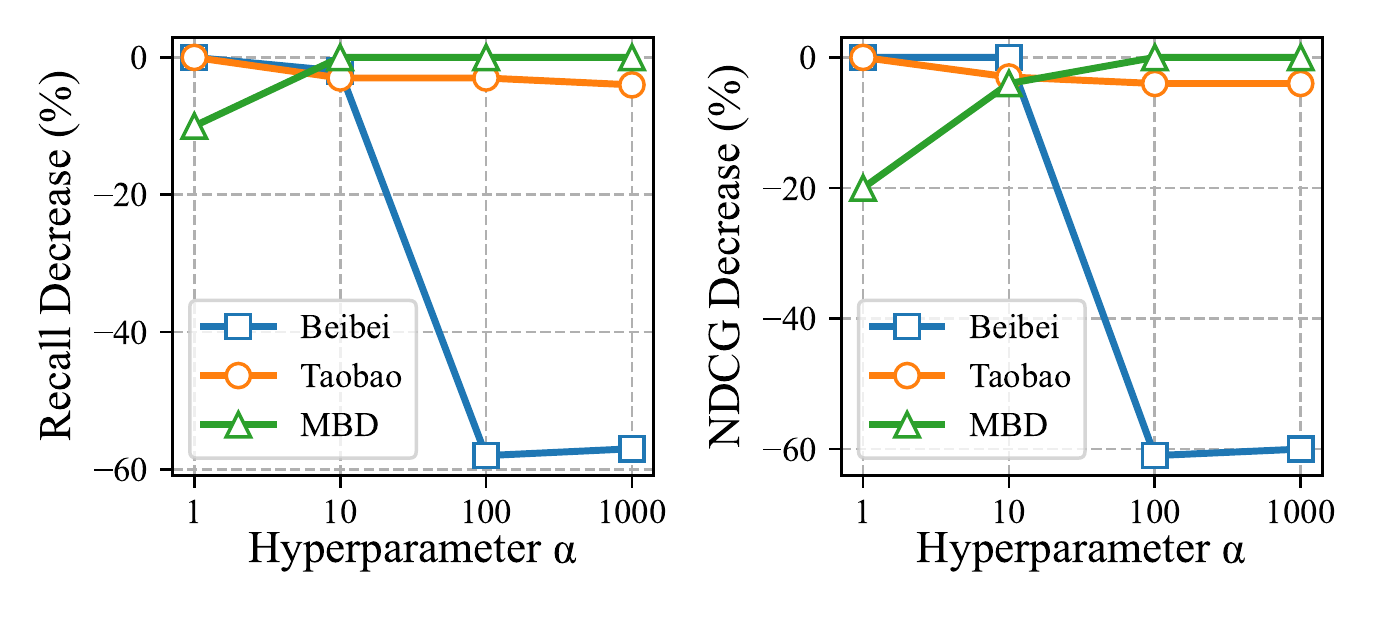}
    % \vspace*{-5mm}
    \caption{Impact of the hyperparameter $\alpha$.}
    \label{fig:alpha}
\end{figure}

Next, we conduct experiments to examine the effect of different parameter settings on MBA.
Figure~\ref{fig:alpha} shows the effect of $\alpha$, which is used to control the weight of the \ac{KL-divergence} in conveying information.
On the Beibei dataset, the performance of MBA is affected when the $\alpha$ is greater than or equal to 100. Thus, when dominated by \ac{KL-divergence}, MBA's performance will be close to that of the pre-trained models.
On the Taobao and MBD datasets, when $\alpha$ is greater than or equal to 100, MBA will gradually converge, with a relatively balanced state between the \ac{KL-divergence} and the expectation term. Under this setting, MBA achieves the best performance.

% Figure~\ref{fig:C1C2} shows the effect of $C_1$ and $C_2$, which are two hyperparameters used in alternative model training.
% Specifically, a large $C_1$ denotes that a negative sample is more likely to be really negative, while a large $C_2$ means that an interacted sample is more likely to be really positive.
% In the three datasets, MBA performs best when $C_1$ and $C_2$ are set to 100.
% When $C_2$ is much larger than $C_1$, the performance of MBA will drop significantly in Taobao and MBD datasets.
% This means that the items which users have interacted with may not be the ones that users really like.
% While in the Beibei dataset, when $C_1$ is much larger than $C_2$, the performance of MBA will drop significantly.
% This means that there are still a large number of potentially interesting items among the un-interacted items.
% \begin{figure}[t]
%     % \setlength{\belowcaptionskip}{-0.75cm}
%     \centering
%     \includegraphics[width=0.46\textwidth]{figures/sigir_C1C2.pdf}
%     \caption{Impact of the hyperparameters $C_1$ and $C_2$.}
%     \label{fig:C1C2}
% \end{figure}
%!TEX root = ../main.tex

\section{Conclusion}
In this work, we have focused on the task of multi-behavior recommendation.
We conjectured that multiple types of behavior from the same user reflect similar underlying user preferences.
To tackle the challenges of the gap between data distributions of different types of behavior and the challenge of behavioral data being noisy and biased, we proposed a learning framework, namely multi-behavior alignment (MBA), which can infer universal user preferences from multiple types of observed behavioral data, while performing data denoising to achieve beneficial knowledge transfer.
Extensive experiments conducted on three real-world datasets showed the effectiveness of the proposed method.

Our method proves the value of mining the universal user preferences from multi-behavior data for the implicit feedback-based recommendation.
However, a limitation of MBA is that  it can only align between two types of behavioral data.
As to our future work, we aim to perform alignment on more types of user behavior.
In addition, we plan to develop ways of conducting more effective and efficient model training.

%\section*{Reproducibility}
\begin{acks}
This research was funded by the Natural Science Foundation of China (62272274,61972234,62072279,62102234,62202271), Meituan, the Natural Science Foundation of Shandong Province (ZR2022QF004), the Key Scientific and Technological Innovation Program of Shandong Province (2019JZZY010129), Shandong University multidisciplinary research and innovation team of young scholars (No.2020QNQT017), the Tencent WeChat Rhino-Bird Focused Research Program (JR-WXG2021411), the Fundamental Research Funds of Shandong University, the Hybrid Intelligence Center, a 10-year program funded by the Dutch Ministry of Education, Culture and Science through the Netherlands Organization for Scientific Research, \url{https://hybrid-intelligence-centre.nl}.

All content represents the opinion of the authors, which is not necessarily shared or endorsed by their respective employers and/or sponsors.
\end{acks}

% \appendix
% \input{sections/appendix}

\clearpage
\bibliographystyle{ACM-Reference-Format}
\balance
\bibliography{references}

\end{sloppypar}
\end{document}